\documentclass[proceedings, preprint]{rmaa}

\usepackage{paralist}

\usepackage{psfrag,color}

\usepackage{natbib}

\SetYear{2009}
\SetConfTitle{The interferometric view of hot stars}

\title{The place of interferometry in massive star multiplicity sudies}

\author{
  H. Sana,\altaffilmark{1,2} 
  and J.-B. Le Bouquin\altaffilmark{1}}

\altaffiltext{1}{European Southern Observatory, Alonso de Cordova 3107,
  Casilla 19001, Santiago 19, Chile (hsana, jlebouqu@eso.org)}
\altaffiltext{2}{Sterrenkundig Instituut 'Anton Pannekoek, Universiteit van Amsterdam, 
  Postbus 94249, 1090 GE Amsterdam, The Netherlands}

\shortauthor{Sana \& Le Bouquin}
\shorttitle{The place of interferometry in massive star multiplicity sudies}

\listofauthors{H. Sana \& J.-B. Le Bouquin}
\indexauthor{Sana, H.}
\indexauthor{Le Bouquin, J.-B.}

\abstract{While it is well known that most massive stars are found to be part of binary
or multiple systems, an accurate characterization of the statistical properties of these multiple 
objects is still lacking. In the present talk, we will review the current status of the
 field, emphasizing the need of using complementarity techniques to cover the large parameter space.
 We will also describe what we think is the place of interferometry in this context.}

\addkeyword{techniques: high angular resolution}
\addkeyword{techniques: interferometric}
\addkeyword{techniques: radial velocities}
\addkeyword{binaries: spectroscopic}
\addkeyword{binaries: general}
\addkeyword{stars: early-type}

\begin{document}
\maketitle

\section{Introduction}
\label{sec:intro}
Despite their importance in modern astrophysics, the massive O-type stars remain incompletely understood. This reflects their actual rareness, and the subsequent large distance at which they are found. As a consequence, even basic parameters such as their mass remain very difficult to accurately measure. 

Fortunately, massive stars are very often found in multiple systems. So far however, only SB2 spectroscopic eclipsing binaries (SB2E) offer reliable constraints on the absolute masses. According to recent reviews \citep{Gie03, SMAW08}, less that 25 direct measurements have been achieved. This is far insufficient to cover a parameter space that spans 80~M$_\odot$ in mass, 20,000~K in temperature, 1.5~dex in luminosity and a factor two (15~R$_\odot$) in radius, not to mention metallicity and rotational velocity.

When trying to characterize the multiplicity properties of massive stars (distribution of orbital parameters, companion properties, etc.), one again faces a very large parameter space. Typical orbital separations span four orders of magnitudes, periods range from a few days to thousands of years, and companions potentially populate the whole mass spectrum. In those conditions, any attempt to explore a significant part of the parameter space needs to take advantage of the complementarity offered by different observational techniques. 

Using adaptive optics, \citet{TBR08} observed about a third of the known galactic O-stars and found companions in 27\%\ of the cases. Using speckle interferometry, \citet{MHG09} targeted most of the galactic O-stars and found companions for 11\%\ of their sample stars. In an extensive review of the literature, the same authors reported that 51\%\ of the O-type objects are actually spectroscopic binaries. Although the coverage of the parameter space is still far from complete and although the observational biases are not uniform, these studies definitely prove that, for massive stars,  binarity is the rule, not the exception.

\begin{figure}[!t]
  \includegraphics[width=\columnwidth]{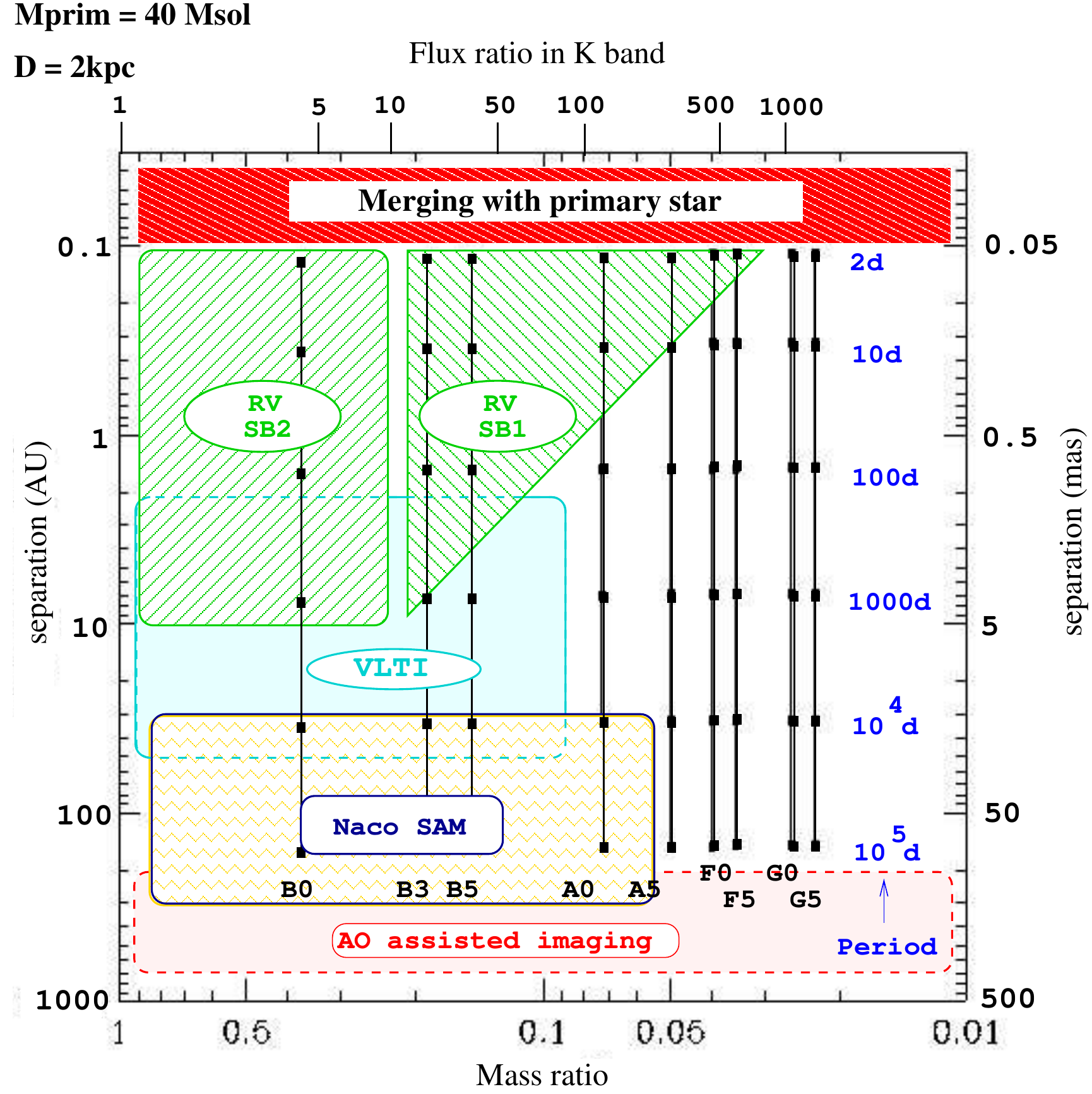}
  \caption{Sketch of the typical parameter space accessible using different observational techniques, with an emphasis on ESO/VLT instrumentation. A main sequence 40~M$_\odot$ primary star has been assumed. Mass-ratio vs.\ physical separation tracks are indicated for various companion types, ranging from B0 ($\sim15$~M$_\odot$) to G5 ($\sim0.9$~M$_\odot$), and various orbital periods (non-eccentric orbits assumed). Separation scale in milli-arcsec (mas, right-hand scale) is computed assuming a distance $D$ of 2~kpc. Primary over secondary flux ratio in the K-band is also given on  top of the graph. }
  \label{fig: param}
\end{figure}

\section{Long-baseline interferometry}
\label{sect: interfero}
Fig.~\ref{fig: param} provide an overview of the typical parameter space of massive binaries and summarises, with an emphasize on ESO/VLT instrumentation, the area of pertinence of different observational techniques. While different approaches are definitely needed to significantly explore the parameter space, these are not equally efficient, nor are they equally demanding in terms of infrastructure and telescope time. Spectroscopy of bright objects is well mastered and relatively cheap. Yet, it is inefficient to detect long period, eccentric binaries. It is further limited to mass-ratio about $q\approx0.2-0.3$ and cannot  retrieve the orbital inclination. It is thus unable to constrain the absolute masses. Adaptive optics only addresses the very large separation, a range where it is difficult to prove the physical bound between the components, especially in relatively crowded field. Speckle interferometry and aperture masking are very efficient in terms of telescope time, and  are thus suitable for blind surveys. Yet, they are also limited to wide separation, with periods in the range 10-100~yr. Long-baseline interferometry, on the other hand, can achieve large flux contrast on separation scale of the order of 1-20~mas. It is however more expensive in terms of infrastructure and telescope time, and is thus best suited to study specific targets. However, long baseline interferometry allows now to cross the gaps between spectroscopic and high-resolution  imaging techniques.

Among the many possible applications, one of the simplest but perhaps one of the most important takes advantage of the increased performances of the VLTI at Paranal. By breaking the $K=7$ magnitude limit, Amber+Finito offers the opportunity to significantly increase the number of O-stars with an accurate mass measurement.  As shown in Fig.~1, SB2 systems with period in the range 200 to 5000~d at 1 to 3~kpc are suitable for both spectroscopy and interferometry. Combining those two techniques allows in principle to recover the orbits in the 3D space, overcoming the uncertainty on the inclination. 


\begin{figure}[!t]
  \includegraphics[width=.49\columnwidth]{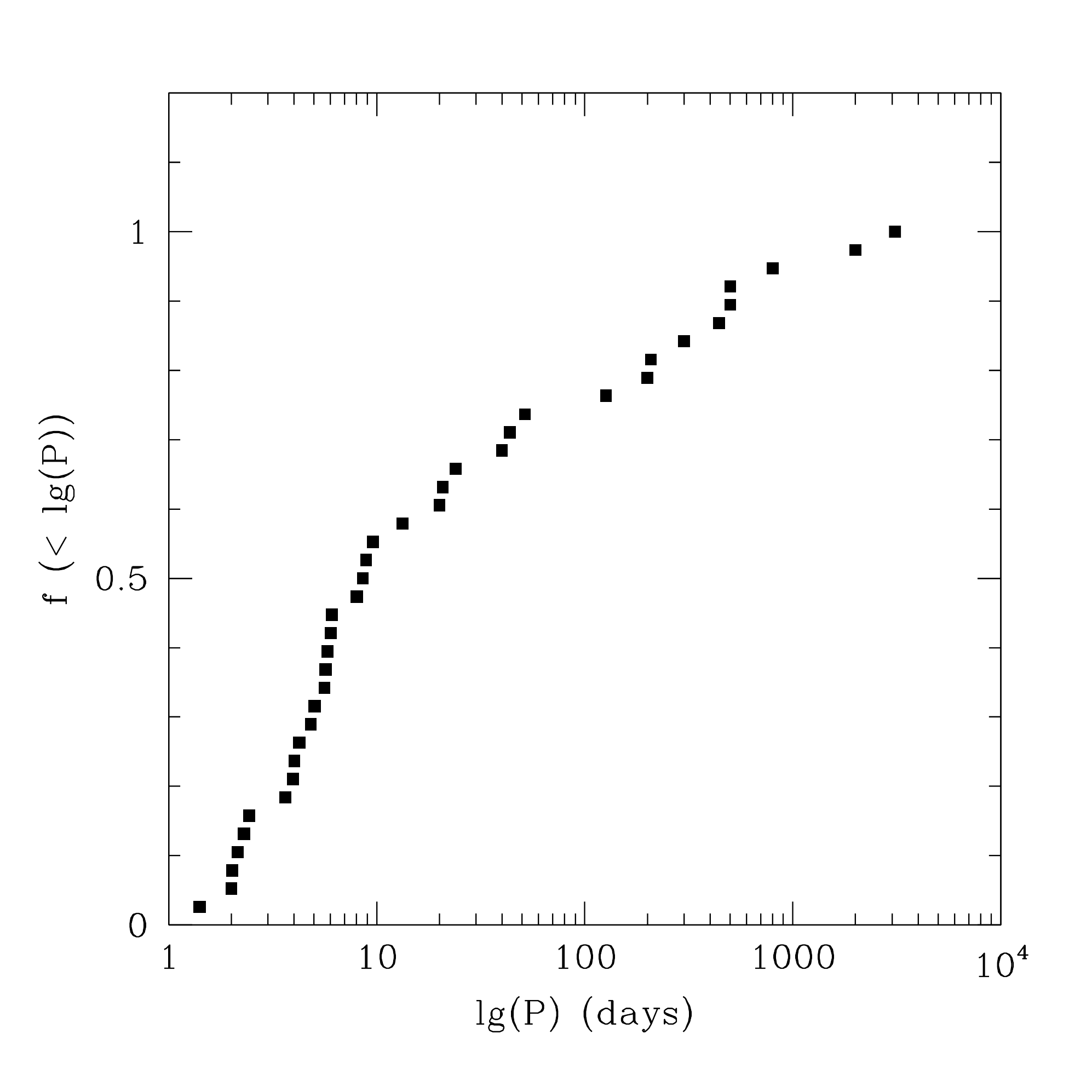}
  \includegraphics[width=.49\columnwidth]{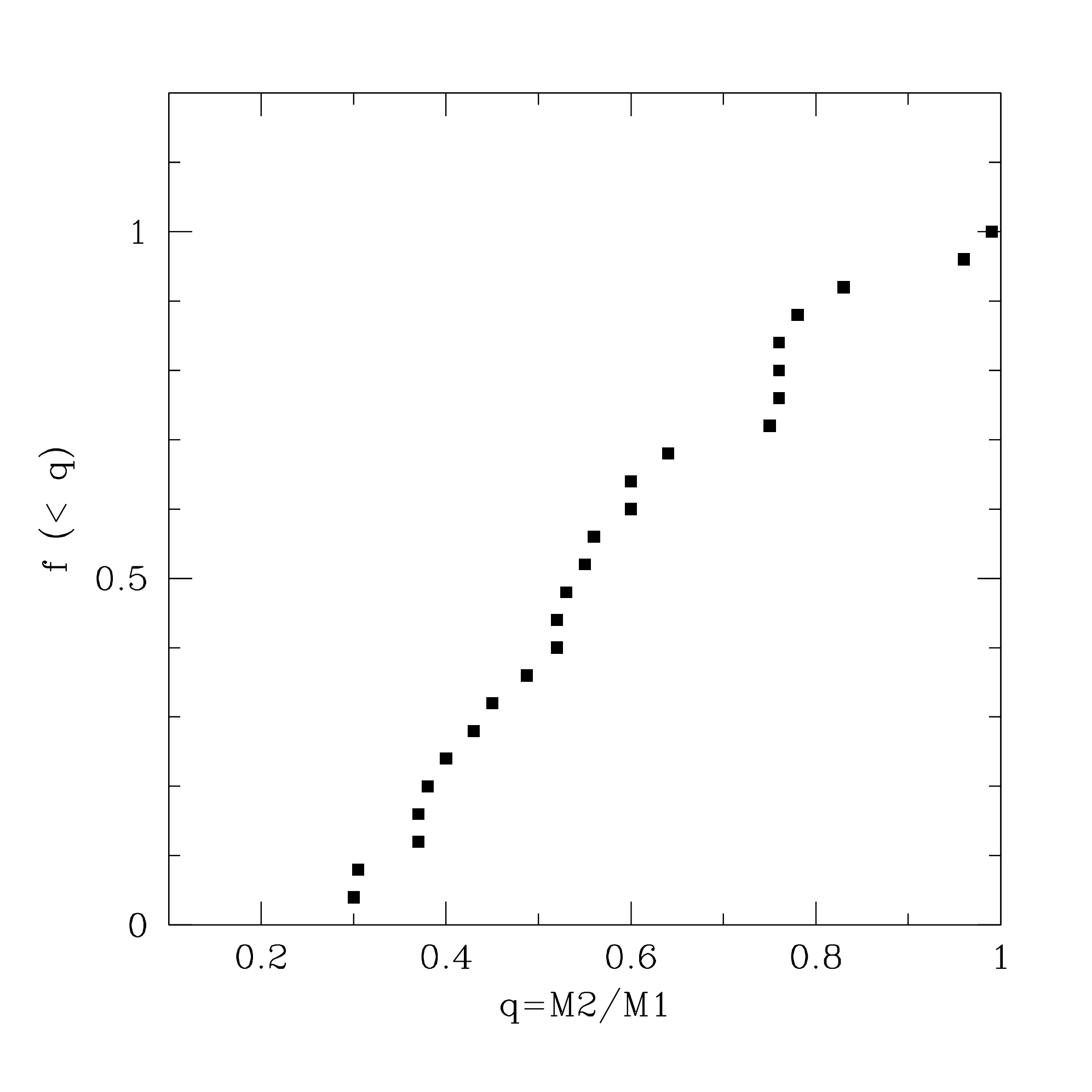}
  \caption{Cumulative distribution of the periods and mass-ratios of the binaries in nearby open clusters. }
  \label{fig: pq}
\end{figure}

\section{Nearby open clusters}
\label{sect: clusters}
The nearby open clusters and OB associations offer a natural selection of potential close-by targets. Fig.~\ref{fig: pq} summarises our current knowledge of the properties of the O+OB binaries in six galactic clusters (IC~1805, IC~2944, NGC~2244, NGC~6231, NGC~6611 and Tr~16) that have been extensively studied by spectroscopy. It reveals that at least 10 objects (25\% of the detected binary population) are located within the parameter space accessible by both spectroscopy and interferometry. This illustrates the fact that the number of potential targets is larger than what is commonly anticipated. The main difficulty resides in finding those binaries whose detection probability, using spectroscopy, is rather low. Yet, long-term monitoring campaigns \citep[e.g., ][]{SGN08} have proved to successfully reveal those systems. Interferometric observations of some of these objects are to be attempted in the coming ESO period. If successful, they will allows us to validate the proposed approach, offering the perspective of new accurate mass measurements of O type stars

\section*{Discussion}
{\bf D.~Baade:} Some young clusters in the Magellanic Clouds are much more massive than in the Galaxy. Can an thing be said about multiplicity of O stars in such environments ?

{\bf H.~Sana:} Certainly. As an example, \citet{BTT09} have showed that 50\% of the massive stars they studied in NGC~2070 in the 30 Dor region were binaries. While a proper assessment of the observational biases is still lacking, their results give a clear indication that the binary fraction in those clusters is likely to be high as well.
 
{\bf H.~Zinnecker:} You briefly alluded to the former suggestion of an anti-correlation of the massive binary fraction and the cluster central density. Could you elaborate a little more on your own view of this claim ?

{\bf H.~Sana:} Such an anti-correlation has been proposed by, e.g., \citet{PGH93} and, later on, supported by \citet{GM01}. The underlying scenario is that binaries in  core of dense clusters are subject to more dynamical interactions than in looser clusters. As a result, dense clusters destroy their binary population and end up with a smaller binary fraction. A number of measurements used by \citeauthor{GM01} to support this scenario have however been significantly revised. As a matter of fact, all the nearby clusters that have been reanalysed since 2001, displays a binary fraction that looks rather uniform within the uncertainties and close to 50\%. My view is thus that the current data do not support such anti-correlation, although more extreme environment should be tested as well.

{\bf D.~Gies:} The presented results on nearby open clusters, and the results from other studies, such as \citet{MHG09}, show that most of the O-type stars are found in high-mass binaries. 
Maybe one should consider that this is a fundamental property \adjustfinalcols  of  the O stars and that this is telling us something about their formation scenario. Maybe binary is the way that massive stars have to circumvent their angular momentum problem. 

{\bf H.~Sana:} I couldn't agree more. Binarity is definitely the rule among O stars and star formation people would need to address the problem into more details at some point. The final product of massive star formation is not a single massive star, but two of them, in a close, short period binary.

\bibliography{/home/hsana/LITERATURE/literature}

\end{document}